\documentclass{pspum-l}

\usepackage{graphicx}

\usepackage{caption}

\newtheorem{theorem}{Theorem}[section]

\theoremstyle{definition}

\theoremstyle{remark}

\numberwithin{equation}{section}

\newcommand{\be}{\begin{equation}}
\newcommand{\ee}{\end{equation}}
\newcommand{\cB}{{\mathcal B}}
\newcommand{\cN}{{\mathcal N}}
\newcommand{\cS}{{\mathcal S}}
\newcommand{\ibar}{{\overline{\imath}}}
\newcommand{\del}{{\partial}}

\newcommand{\Cbar}{{\overline{C}}}
\newcommand{\tbar}{{\overline{t}}}
\newcommand{\psibar}{{\overline{\psi}}}
\newcommand{\bP}{{\mathbb P}}

\begin{document}

\title[Resurgence]{Resurgence and Topological Strings}


\author{M. Vonk}
\address{Institute for Theoretical Physics, University of Amsterdam, Science Park 904, Postbus 94485, 1090 GL Amsterdam, The Netherlands}
\email{m.l.vonk@uva.nl}
\thanks{The research of the author was supported by the European Research Council Advanced Grant EMERGRAV}



\date{}

\begin{abstract}
The mathematical idea of resurgence allows one to obtain nonperturbative information from the large--order behavior of perturbative expansions. This idea can be very fruitful in physics applications, in particular if one does not have access to such nonperturbative information from first principles. An important example is topological string theory, which is a priori only defined as an asymptotic perturbative expansion in the coupling constant $g_s$. We show how the idea of resurgence can be combined with the holomorphic anomaly equation to extend the perturbative definition of the topological string and obtain, in a model--independent way, a large amount of information about its nonperturbative structure.
\end{abstract}

\maketitle


\section{Introduction}
The vast majority of calculational problems in physics are impossible to solve exactly. For this reason, it is important to have good approximation techniques at one's disposal. One such technique is the perturbative approach: one identifies a (preferably small) parameter $x$ in the problem, such that the problem can be solved exactly in the special case where $x=0$. Then, one tries to construct the full solution $f(x)$ to the problem order by order in a perturbative expansion:
\be
 f(x) = \sum_{n=0}^\infty a_n x^n.
\ee
Of course, it will in general not be possible to find a closed form for all the coefficients $a_n$ (that would essentially amount to finding an exact solution to the problem), but often one can calculate the individual coefficients one by one, up to arbitrarily high $n$. One may then calculate a partial sum of the form
\be
 f_N(x) = \sum_{n=0}^N a_n x^n
\ee
and, for large $N$, view such a sum as an approximation to the true answer $f(x)$.

There are two well--known problems that may arise in this approach. The first one is that the partial sums $f_N(x)$ may not converge when one takes the limit $N \to \infty$. The canonical example where this happens is the Taylor expansion
\be
 \frac{1}{1-x} \approx 1 + x + x^2 + x^3 + \ldots
\ee
Here, the partial sums of the right hand side diverge when $|x|>1$, even though for any $x \neq 1$, the left hand side is well--defined. This example still has a finite radius of convergence, but there are other examples where the perturbation series is {\em asymptotic}: even though the partial sums become better and better approximations as $x \to 0$, for any $x \neq 0$ they diverge when $N \to \infty$. A famous example is Stirling's approximation to the logarithm of the Gamma function:
\be
 \log \Gamma(z) - \frac12 \log(2 \pi) - \left( z - \frac12 \right) \log z + z \approx \frac{1}{12 z} - \frac{1}{360 z^3} + \frac{1}{1260 z^5} - \ldots
\ee
where now $z^{-1}$ is the small parameter, and we have moved all terms that do not involve positive powers of this parameter to the left hand side. One can show that the coefficients of $z^{1-2n}$ on the right hand side grow like $\frac{(2n-2)!}{(2\pi)^{2n}}$. From this factorial growth of the coefficients, one then easily shows that the partial sums diverge for any nonzero value of $z^{-1}$. This factorial growth of perturbative coefficients, and the resulting asymptotic behavior, is ubiquitous in quantum mechanics and quantum field theory.

A second problem that often arises when one uses perturbative methods is that some nontrivial functions have a vanishing perturbative expansion. Here, the canonical example is
\be
 f(x) = e^{-\frac{1}{x^2}}
\ee
which is a well--defined and smooth function on the real axis, but which has a vanishing Taylor series around $x=0$. More generically, instanton and soliton effects in physics often cannot be ``seen'' in perturbation theory.

An important observation is that the above two problems are not at all independent, as can be seen e.g.\ from Borel resummation. In section \ref{sec:resurgence}, we will review how this comes about, and how this relation between asymptotic perturbative series and nonperturbative effects takes its full form in the theory of resurgence. The idea of resurgence has great potential in physical applications, since it allows us to obtain nonperturbative information from a purely perturbative expansion. In this contribution, based on the work \cite{Santamaria:2013rua,Couso-Santamaria:2014iia} with R.~Couso-Santamar\'ia, J.~D.~Edelstein and R.~Schiappa, we want to work out this idea for the example of topological strings. This example is particularly interesting, as topological strings are a priori only {\em defined} as an asymptotic perturbative expansion. Finding a generic way to extend their partition sums into full, nonperturbative functions, is therefore a very interesting open question. In section \ref{sec:ha}, we explain how the holomorphic anomaly equation provides a window into this problem. In section \ref{sec:resha}, we then present some explicit results for the example of B--model toplogical strings on local $\bP^2$. We end with a conclusion and outlook in section \ref{sec:conclusion}.

\section{Resurgence}
\label{sec:resurgence}
In this section, we very briefly review some of the basic ideas of resurgence. The reader is referred to \cite{Aniceto:2011nu} and references therein for a more thorough introduction.

\subsection{Borel transforms and asymptotic behavior of coefficients}
If a quantity $f(x)$ has a divergent perturbation series whose coefficients grow as
\be
 \label{eq:coeffgrowth}
 a_n \sim \frac{n!}{A^n}
\ee
for some (generically complex) $A$, then there is a simple trick due to Borel that often allows one to find a well--defined function for which the given series is an asymptotic perturbative approximation. One defines the {\em Borel transform} $\cB[f]$ as the formal power series
\be
 \cB[f](s) = \sum_{n=0}^{\infty} \frac{a_n}{n!} s^n.
\ee
This turns the asymptotic series for $f$ into a new series which has a singularity at $s=A$ but is convergent when $|s|<A$. If we now assume that $A$ is not a positive, real number and that $\cB[f](s)$ can be analytically continued to the positive real axis, then one easily checks that
\be
 \label{eq:invbor}
 \cS_0 f(x) \equiv \int_{s=0}^\infty \cB[f](sx) \; e^{-s} ds
\ee
gives back a function $\cS_0 f(x)$, called the {\em Borel resummation} of $f(x)$, which has the same asymptotic expansion as the original function. Thus, this Laplace--type transform can be thought of as a formal inverse of the Borel transform.

Of course, when $A$ is real and positive, the above procedure does not work, since for positive $x$ the integral in (\ref{eq:invbor}) runs into the singularity of $\cB[f]$. One may define two alternative Borel resummations, $\cS_{\pm} f(x)$, by using integration contours in the complex plane which circumvent the singularity either above or below. Using Cauchy's theorem, one then sees that the difference between those two resummations is of the order
\be
 (\cS_- - \cS_+) f(x) \approx e^{-A/x}.
\ee
This result is perhaps not too surprising: the difference between the two Borel resummations --- which each have the same asymptotic expansion as the original function --- is a function of the ``instanton type'', which itself has a vanishing perturbative expansion. This is the relation between asymptotic series and nonperturbative functions that we alluded to in the introduction.

Often, for example for reasons of reality (see e.g.\ \cite{Aniceto:2013fka}), one can show that neither of the two Borel resummations gives back the original function $f(x)$, but that the true function $f(x)$ lies ``in the middle'', in the sense that
\begin{eqnarray}
 f(x) & = & \cS_+ f(x) + \frac12 e^{-A/x} \left( \ldots \right) \nonumber \\
 & = & \cS_- f(x) - \frac12 e^{-A/x} \left( \ldots \right)
 \label{eq:fresum}
\end{eqnarray}
where the dots indicate the resummation of a further expansion that we will make more precise in the next subsection.

A crucial observation to make about this discussion is that the coefficient $A$ (usually called the ``instanton action'') which can be read off from the asymptotic growth of the {\em perturbative} coefficients in (\ref{eq:coeffgrowth}), also appears in the {\em nonperturbative} contributions in (\ref{eq:fresum}). Somehow, the perturbative solution to our problem already knows something about its nonperturbative completion. As we will see, resurgence makes this statement precise, and extends it immensely.

\subsection{Transseries}
As we have seen in the previous subsection, a perturbative power series contains some information (like the instanton action $A$) about the nonperturbative contributions to the function one wants to describe. At the same time, this information is not encoded in a very straightforward way. To make the nonperturbative content of a function more transparent, it turns out to be very useful to describe it using a {\em transseries}. A simple example of a transseries is an expression of the form
\be
 \label{eq:simplets}
 f(x) \approx \sum_{k=0}^\infty \sum_{n=0}^\infty a^{(k)}_n e^{-kA/x} x^n.
\ee
In this case, the transseries is a formal expansion in the variable $x$ and in a ``nonperturbative building block'' $e^{-A/x}$. More general transseries may have many more of such building blocks: they can be expansions in several different instanton factors, $e^{-A_i/x}$, expansions in $\log(x)$ or other nonanalytic functions of $x$, etc. We will not discuss the general theory of transseries here --- the reader can find several good references in \cite{Aniceto:2011nu}. For most of this paper, formal transseries of the above form will be sufficient.

Transseries solutions to physics problems often arise in a very natural way. For example, if the problem is described by a differential or finite difference equation, one can often simply insert an ansatz of the form (\ref{eq:simplets}) into this equation and solve it order by order to find the coeffients $a^{(k)}_n$ and the instanton action $A$ --- in the same way that one would construct a formal power series solution. In fact, this is exactly what we will do in this paper.

Of course, given a transseries, one may again ask how to obtain an actual {\em function} from it. To this end, let us write the transseries expansion (\ref{eq:simplets}) as
\be
 \label{eq:tscomp}
 f(x) \approx \sum_{k=0}^\infty e^{-kA/x} \Phi^{(k)}(x) \qquad \mbox{with} \qquad \Phi^{(k)}(x) = \sum_{n=0}^\infty a^{(k)}_n x^n.
\ee
In general, all of the pertrubative series $\Phi^{(k)}(x)$ may be asymptotic, divergent series. However, as was the case for ordinary power series, one can turn a formal transseries into an actual function by choosing an integration contour and Borel resumming each sector. That is, one extends the definition of Borel resummation in the natural way to be
\be
 \label{eq:borelts}
 \cS_+ f(x) = \sum_{k=0}^\infty e^{-kA/x} \int_{s=0^+}^\infty \cB \left[ \Phi^{(k)} \right](sx) \; e^{-s} ds
\ee
where we have chosen the $+$--contour to be specific.

Note that expressions of the form (\ref{eq:tscomp}) appear very naturally in physics. In many problems, e.g.\ in quantum field theory, one expands a solution around a trivial ``vacuum'' background. However, there are in general other, nonperturbative backgrounds, such as instantons and solitons, that one wants to take into account as well. Each of these backgrounds is suppressed by a nonperturbative factor, often of the form $e^{-A/x}$, and one can construct a new, perturbative solution around such a background. One then wants to ``add up'' all of the different nonperturbative sectors into a single solution to the problem. The prescription (\ref{eq:borelts}) makes this procedure mathematically precise.

\subsection{Resurgence}
Even though we can formally write down any transseries we like, the transseries solutions that arise from physical or mathematical problems usually have a lot of extra structure. We saw an example of this in equation (\ref{eq:coeffgrowth}) and (\ref{eq:fresum}) where, essentially just from the requirement of analyticity, we derived that the instanton action $A$ could be read off from the large--order behavior of the perturbative coefficients $a^{(k)}_n$. The theory of resurgence, first developed by J.~\'Ecalle in \cite{Ecalle:1981}, extends this example immensely, and makes the resulting structure very precise.

In \cite{Ecalle:1981}, a class of functions called {\em resurgent functions} is defined. The defining property is that the Borel transform of a resurgent function only has a discrete set of singularities and any analytic continuation along a path avoiding these singularities can be defined. For our purposes, the precise definition is not very relevant; all that matters is that most of the functions one encounters in (toy model) physical examples belong to the class of resurgent functions.

One can show that for resurgent functions, not only the instanton action $A$ can be derived from the perturbative coefficients $a^{(0)}_n$, but in fact {\em all} coefficients $a^{(k)}_n$ in {\em all} other instanton sectors $\Phi^{(k)}$ can be obtained from the perturbative sector. This is the origin of the name ``resurgence'': the instanton sectors can be ``resurrected'' from the perturbative sector alone. Actually, the choice of the perturbative ``vacuum'' sector is somewhat arbitrary: one could also reconstruct the vacuum sector (and all other sectors) from an arbitrary given instanton sector.

The way to obtain this nonperturbative information from perturbative information is essentially through a huge generalization of (\ref{eq:coeffgrowth}). The details of this depend on the problem at hand (and the derivation requires several new mathematical techniques that are explained in \cite{Ecalle:1981}), but in the end it turns out that one can derive large--order relations which are schematically of the form
\be
 a^{(0)}_n \sim \sum_{k=1}^{\infty} \frac{S_1^k}{2\pi i} \sum_{m=0}^{\infty} \frac{\Gamma(n-m)}{(kA)^{n-m}} a^{(k)}_m.
\ee
Here, $S_1$ is an unknown problem--dependent constant known as the {\em Stokes constant}. We stress that the above expression is just schematic: in actual computations (see e.g.\ \cite{Aniceto:2011nu}), there may be many different instanton actions $A_i$, many different Stokes constants, and arguments of the form $n-m$ may be shifted by problem--dependent constants. However, the general structure of these large--order relations and the way they can be used is always the same. Note for example that taking the leading ($k=1, m=0$) contribution in the above equation, we recover the fact that the leading growth of the perturbative coefficients $a^{(0)}_n$ is determined by $A$ as in (\ref{eq:coeffgrowth}). Taking higher $m$ terms into account, we see that $\frac{1}{m}$--corrections to this leading growth determine the one--instanton coefficients $a^{(1)}_n$. Then from the $k=2$ terms, we see that nonperturbative corrections of order $(2)^{-m}$ to this growth determine the two--instanton coefficients $a^{(2)}_n$, and so on.

The use of large--order relations in physical applications is twofold. First of all, there are problems where the perturbative sector can be calculated, but where the nonperturbative contributions are unknown --- either due to computational difficulties, or for more fundamental reasons. In fact, the example of topological strings that we will discuss is a case where the theory is only defined perturbatively, and there is no generally defined nonperturbative completion. In such cases, one may use large--order relations to calculate nonperturbative contributions from consistency conditions alone.

Of course, in doing so, one {\em assumes} that the function one is investigating belongs to the class of resurgent functions. This can in general not be proven, but here, the large--order relation is again useful. If one can obtain conjectured nonperturbative contributions by other means --- for example by plugging a transseries ansatz into a differential or finite difference equation --- one can then use the large--order relation to {\em test} whether the function indeed has resurgent behavior. One may perform these tests for several nonperturbative contributions to gain confidence in the resurgent properties, and then use the large--order relation to calculate further nonperturbative terms. In what follows, we will illustrate this approach using the example of the topological string.

\section{The holomorphic anomaly}
\label{sec:ha}
There are several equivalent ways to define topological string theories and their partition functions. Physically, a topological string theory be obtained by twisting a two--dimensional $\cN=2$ supersymmetric field theory and coupling the resulting theory to two--dimensional gravity. Mathematically, topological string theory partition functions can be defined as generating functions of Gromov--Witten invariants, or they can be obtained from studying the complex structure deformations of Calabi--Yau manifolds. It would go too far to review any of these definitions here; we refer the interested reader to the many available reviews of the topic, such as the extensive book \cite{Hori:2003ic}.

For our purposes, all that matters is that all of these definitions depend on a parameter $g_s$, called the topological string coupling constant, and that they all lead to a partition function which is a perturbative expansion in $g_s$. More precisely, the free energy (which is the logarithm of the partition function) is an expansion of the form
\be
 \label{eq:tsts}
 F(t; g_s) \approx \sum_{g=0}^{\infty} F_g(t) \; g_s^{2g-2}.
\ee
Here, we denoted any additional parameters that the problem may have by $t$; these can for example take the form of couplings in the two--dimensional field theory, or of moduli of the Calabi--Yau manifold.

Calculating the coefficients $F_g(t)$ from one of the definitions of the topological string theory is often very complicated, especially if one wants to go beyond the first few values of $g$. Fortunately, in the incarnation where the topological string is defined from the complex structure deformations of Calabi--Yau manifolds (the so--called B--model), a shortcut was found in \cite{Bershadsky:1993ta,Bershadsky:1993cx}. It turns out that the $F_g(t)$ are almost holomorphic in the parameters $t^i$, where the ``almost'' means that there is a recursion relation, called the {\em holomorphic anomaly equation}, of the form
\be
 \label{eq:ha}
 \del_{\ibar} F_g = \frac12 {\Cbar_{\ibar}}^{jk} \left( D_j D_k F_{g-1} + \sum_{h=1}^{g-1} D_j F_{g-h} D_k F_h \right).
\ee
The definition of the coefficients ${\Cbar_{\ibar}}^{jk}$ and of the covariant derivatives $D_i$ can be found e.g.\ in \cite{Hori:2003ic}. The crucial point is that the derivative with respect to the anti--holomorphic variable $\tbar^\ibar$ depends {\em only} on the $F_h$ with $h<g$. This allows one to determine the $F_g$ recursively, up to an integration constant which is purely holomorphic in the $t^i$. It turns out that this integration constant can often be determined exactly from the behavior of $F(t)$ at special points in the moduli space; see e.g.\ \cite{Haghighat:2008gw}. This so--called ``direct integration'' method is very efficient, and can be used to compute the $F_g$ up to high values of $g$.

The direct integration technique allows one to construct a formal, perturbative expression for the topological string theory free energy, but the power series one constructs in this way are asymptotic and diverge. This is a serious problem, even more so because, as we have mentioned, the usual definitions of the topological string theory are {\em only} perturbative in nature. To obtain an actual function $F(t)$, we need to somehow identify the nonperturbative contributions and construct a nonperturbative completion of the asymptotic series. While this has been done in specific instances (usually through the use of dualities), a generic procedure does not exist. Clearly, the theory of resurgence can be a major asset to fill this gap.

Thus, one would first of all like to extend the power series solution (\ref{eq:tsts}) to a transseries solution. An immediate problem presents itself, as the holomorphic anomaly equation (\ref{eq:ha}) is an equation for the {\em individual coefficients} $F_g(t)$, not for the full $F(t)$. Thus, we cannot simply insert a transseries ansatz into this equation. Fortunately, as was already pointed out in \cite{Bershadsky:1993cx}, it is not too hard to find an equation that the full partition function $Z = e^{F(t)}$ satisfies. Roughly, $Z$ satisfies a heat kernel equation of the form
\be
 \label{eq:haint}
 \left( \del_{\ibar} - \frac12 g_s^2 {\Cbar_{\ibar}}^{jk} D_j D_k\right) Z = 0.
\ee
By ``roughly'', we mean that some additional terms must be included in the equation to correct the resulting anomaly equations at low--lying $g$. The details of this, and the precise equation that replaces (\ref{eq:haint}), can be found in our paper \cite{Santamaria:2013rua}. In what follows, we will simply refer to the above equation, but the reader should keep in mind that it is only a schematic representation of the true holomorphic anomaly equation for $Z(t)$.

Now, the plan of attack is clear: one can make a transseries ansatz for $F(t)$, plug $Z = e^F$ into the above equation, and recursively solve for all nonperturbative coefficients. As was the case for the ordinary power series solution, this requires the fixing of holomorphic integration constants at every order, which can be done using boundary behavior of $F(t)$ at special points in moduli space. Thus, we can describe this procedure as ``nonperturbative direct integration''.

Following this procedure, we find a conjectural form of the nonperturbative topological string free energy expressed as a transseries. One can then compare these results to existing conjectures for nonperturbative topological strings, and test, as described in the previous section, whether the $F(t)$ one finds is indeed a resurgent function. We have carried this out for general models in \cite{Santamaria:2013rua}, and for the specific example of B--model topological strings on local $\bP^2$ in \cite{Couso-Santamaria:2014iia}. In the next section, we report some of the most interesting results from these papers.

\section{Resurgence and the holomorphic anomaly}
\label{sec:resha}
We will present the results of the procedure outlined above in a mostly graphical manner. For the analytical derivations and numerics behind the images, and a much more detailed description, we refer the reader to \cite{Santamaria:2013rua,Couso-Santamaria:2014iia}. The figures in this section are originally in full--color; the reader can find the colored version of the images in the online version of this paper.

\begin{figure}[t]
\centering
\begin{minipage}{.5\textwidth}
  \centering
  \includegraphics[height=.5\linewidth]{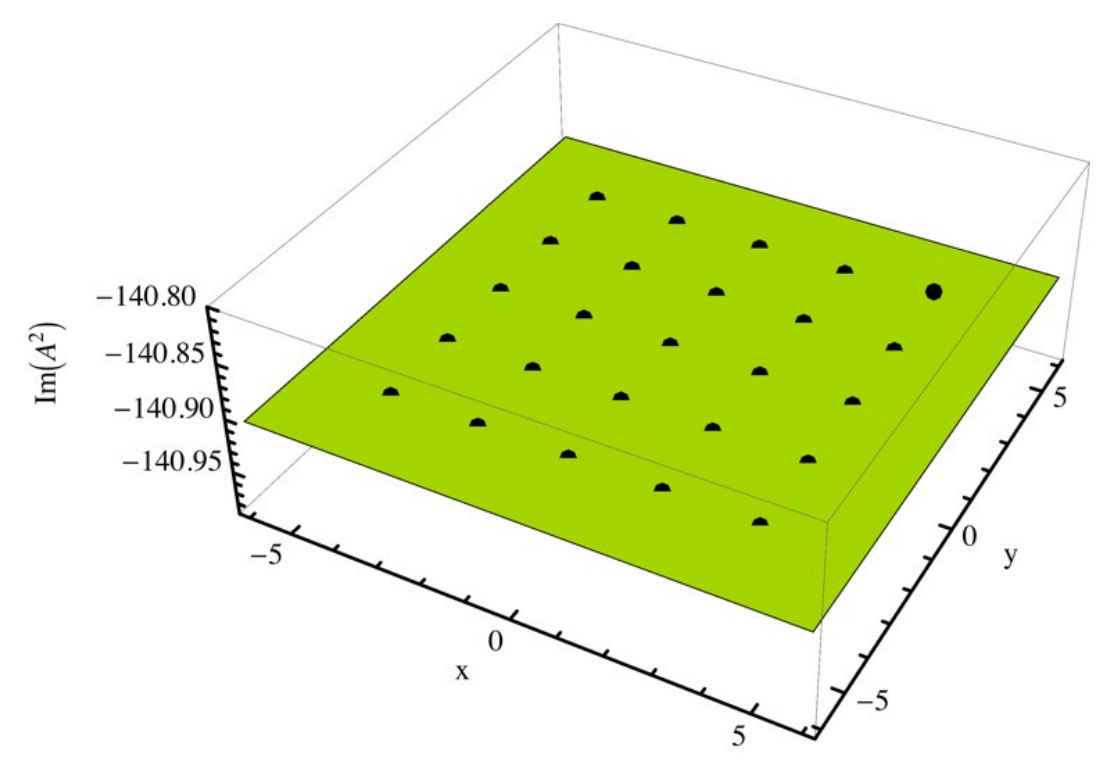}
  \captionsetup{width=5.5cm}
  \captionof{figure}{The instanton action is independent of the anti-holomorphic modulus.}
  \label{fig:1}
\end{minipage}%
\begin{minipage}{.5\textwidth}
  \centering
  \includegraphics[height=.5\linewidth]{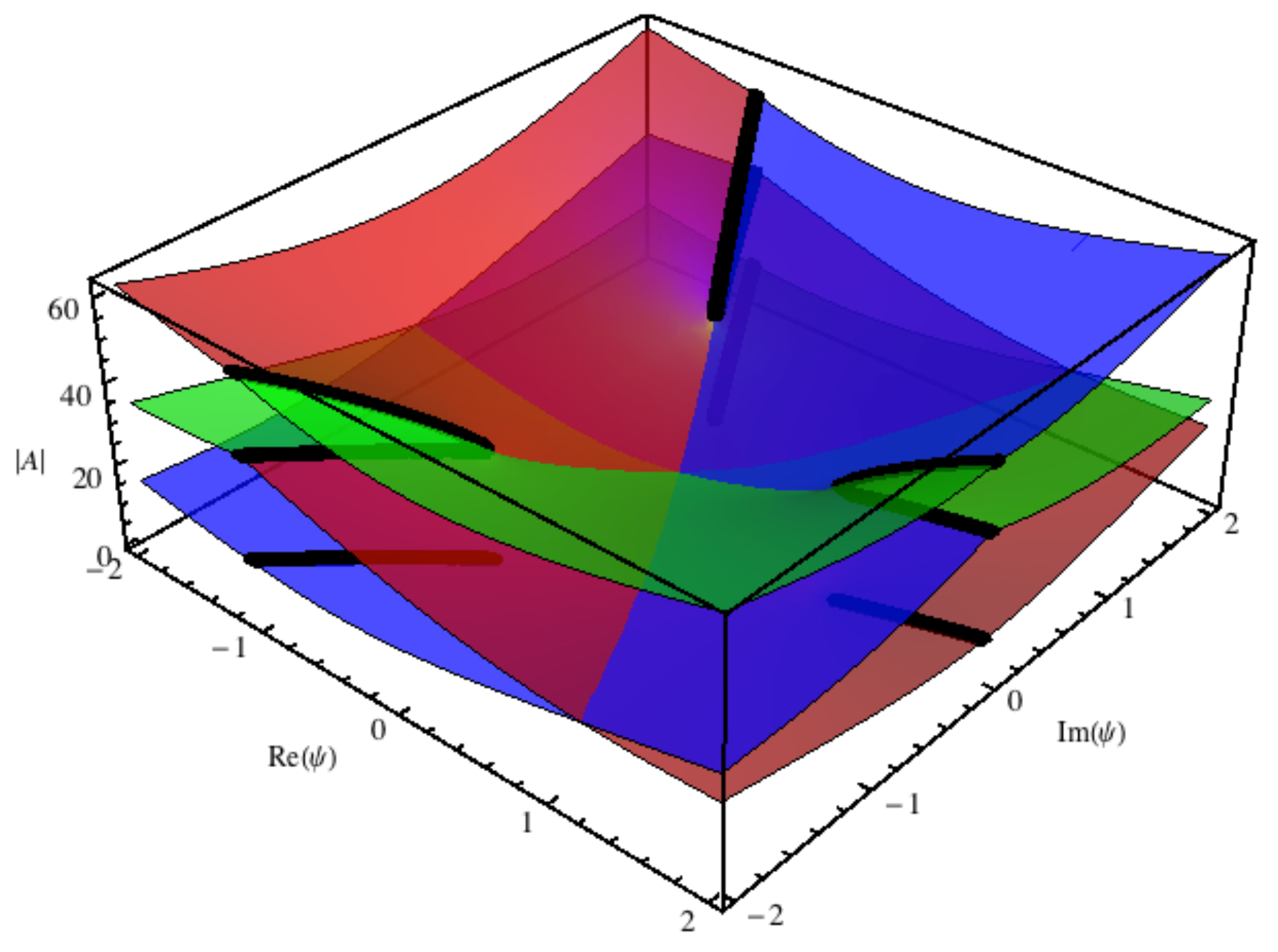}
  \captionsetup{width=5.5cm}
  \captionof{figure}{Local $\bP^2$ has three different conifold instanton actions.\\}
  \label{fig:2}
\end{minipage}
\end{figure}

The most basic prediction of resurgence is that the instanton actions $A$, calculated using a transseries ansatz for the equation (\ref{eq:haint}), can also be obtained from the large--order behavior of the perturbative $F_g(t)$ using the analogue of (\ref{eq:coeffgrowth}). We have found that for topological strings, this is indeed the case. Perhaps more interestingly, using resurgence one can show on general ground that the $\tbar$--dependence disappears in the large $g$ limit, and that the instanton actions are all {\em purely holomorphic} in $t$. This can also be shown numerically in explicit examples. In figure \ref{fig:1}, for example, we see in the local $\bP^2$ example that an instanton action (of which for technical reasons we plot the imaginary part of the square) is independent of the value of an anti--holomorphic modulus $\psibar = x + i y$. This result is consistent with a conjecture made in \cite{Drukker:2011zy}, which states that the instanton actions of topological string theories quite generally are given by periods of holomorphic three--forms --- which naturally depend in a holomorphic manner on the moduli.

Of course, for a generic topological string theory, we expect to find more than one instanton action. In fact, in the case of local $\bP^2$, one easily finds three different conifold periods which can all appear as instanton actions. (There is in fact a fourth period associated to the large--volume limit of local $\bP^2$ which we will ignore in this paper.) In figure \ref{fig:2}, we plot the absolute value of those three periods as functions of the modulus $\psi$ of the model. At each point in moduli space, one can check that the large--order behavior of the $F_g$ is indeed determined by the period $A_i$ with smallest absolute value.

\begin{figure}[t]
  \centering
  \includegraphics[height=.5\linewidth]{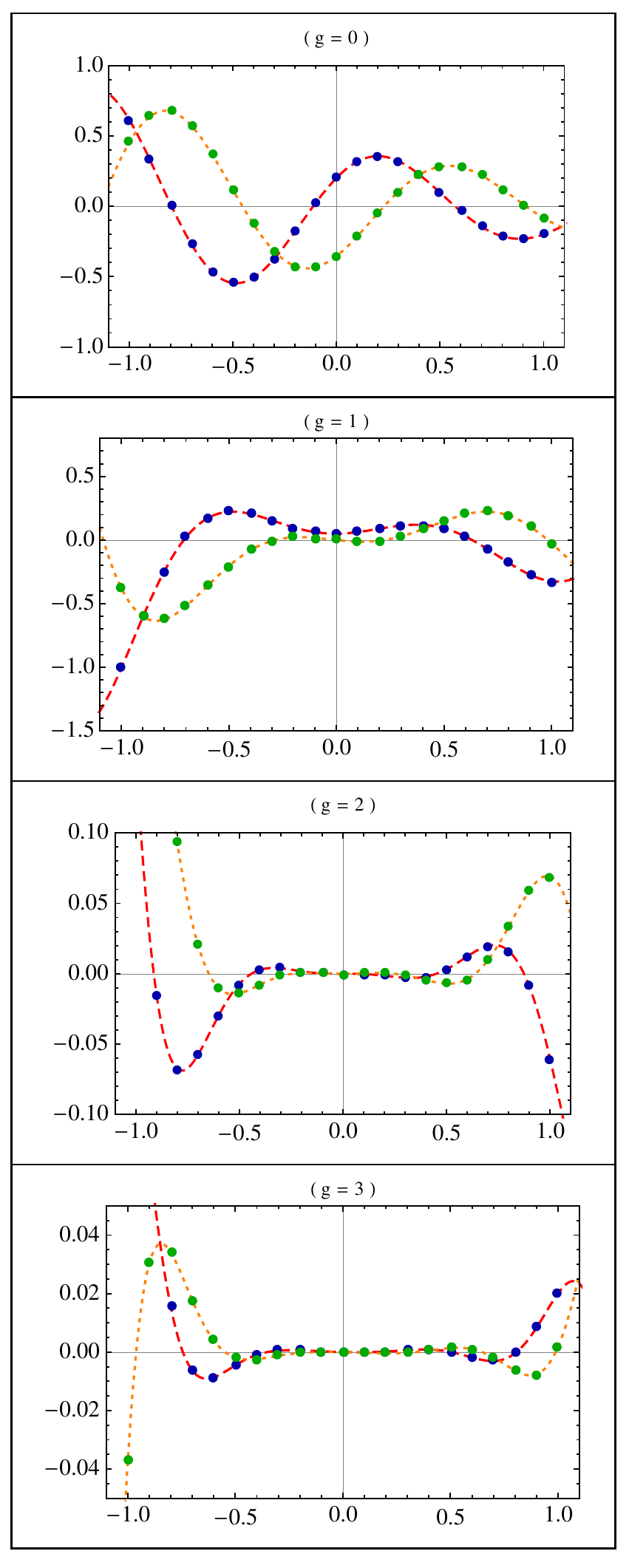}
  \includegraphics[height=.5\linewidth]{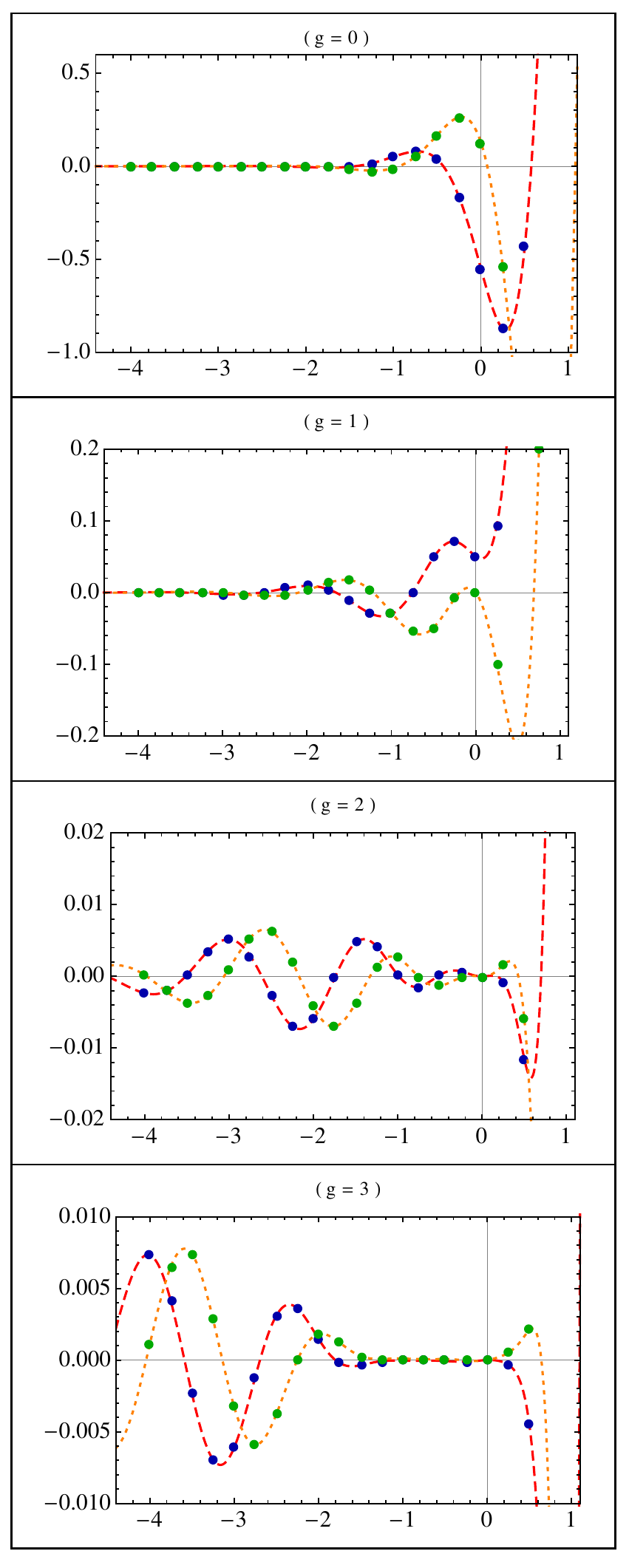}
  \includegraphics[height=.5\linewidth]{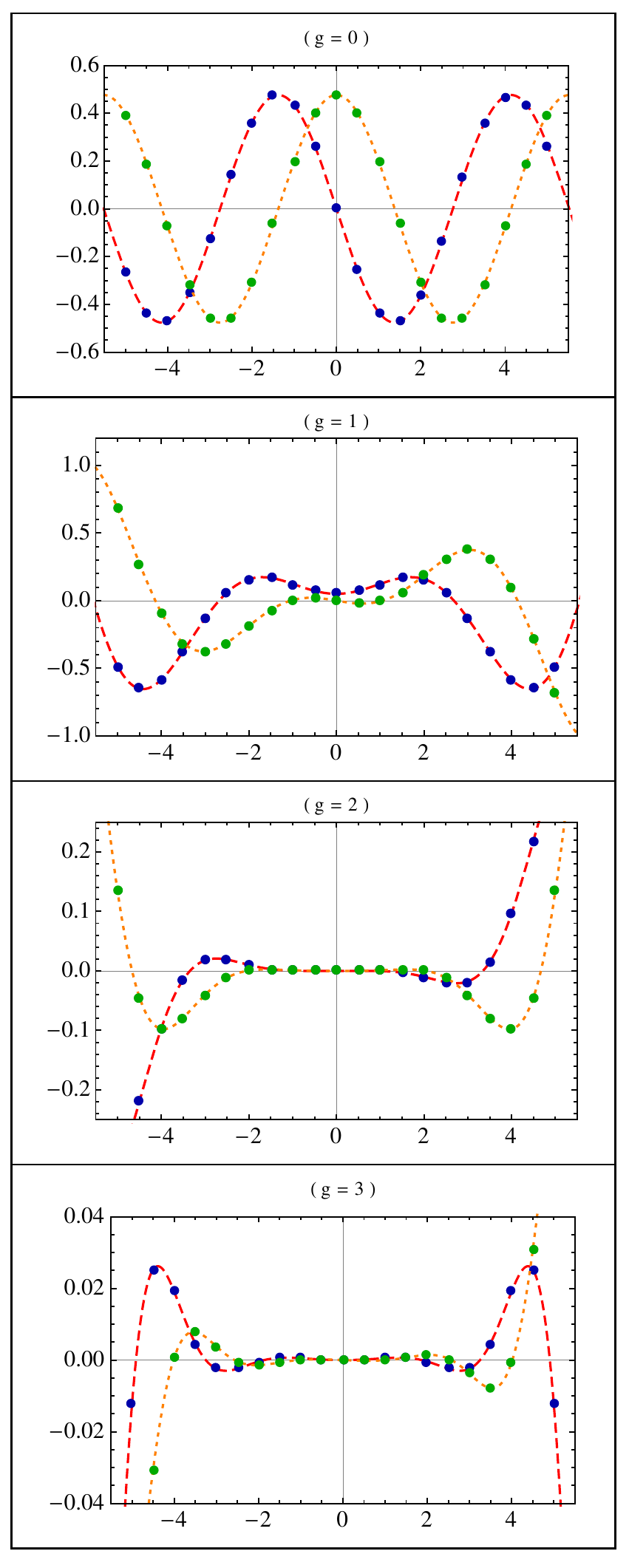}
  \captionof{figure}{Comparison between transseries results and large--order results for several nonperturbative coefficients.}
  \label{fig:3}
\end{figure}

Next, one can use a transseries ansatz and (\ref{eq:haint}) to calculate nonperturbative coefficients $F^{(k)}_g$ in several instanton sectors. One can then check whether these coefficients match the large--order behavior of the perturbative $F_g$. In figure \ref{fig:3}, we show these tests for three different values of the modulus (from left to right) for the first four leading coefficients $F^{(1)}_g$ (from top to bottom). The dependence on the anti--holomorphic modulus $\psibar$ is plotted in the figures. The lines show the predictions (real and imaginary part) from the transseries ansatz; the dots show the results from the large--order behavior of $F_g$. We see that the dots match the lines perfectly, meaning that all of this nonperturbative information is indeed captured by the perturbative $F_g$. As an interesting side remark: we see that for holomorphic moduli (0 on the horizontal axis in the plots) all $F^{(1)}_{g\geq1}$ vanish. This is consistent with the fact that in this limit, from the direct integration procedure one expects to find back the exact conifold results. In \cite{Pasquetti:2009jg}, these coefficients were indeed shown to vanish.

\begin{figure}[t]
\centering
\begin{minipage}{.5\textwidth}
  \centering
  \includegraphics[height=.55\linewidth]{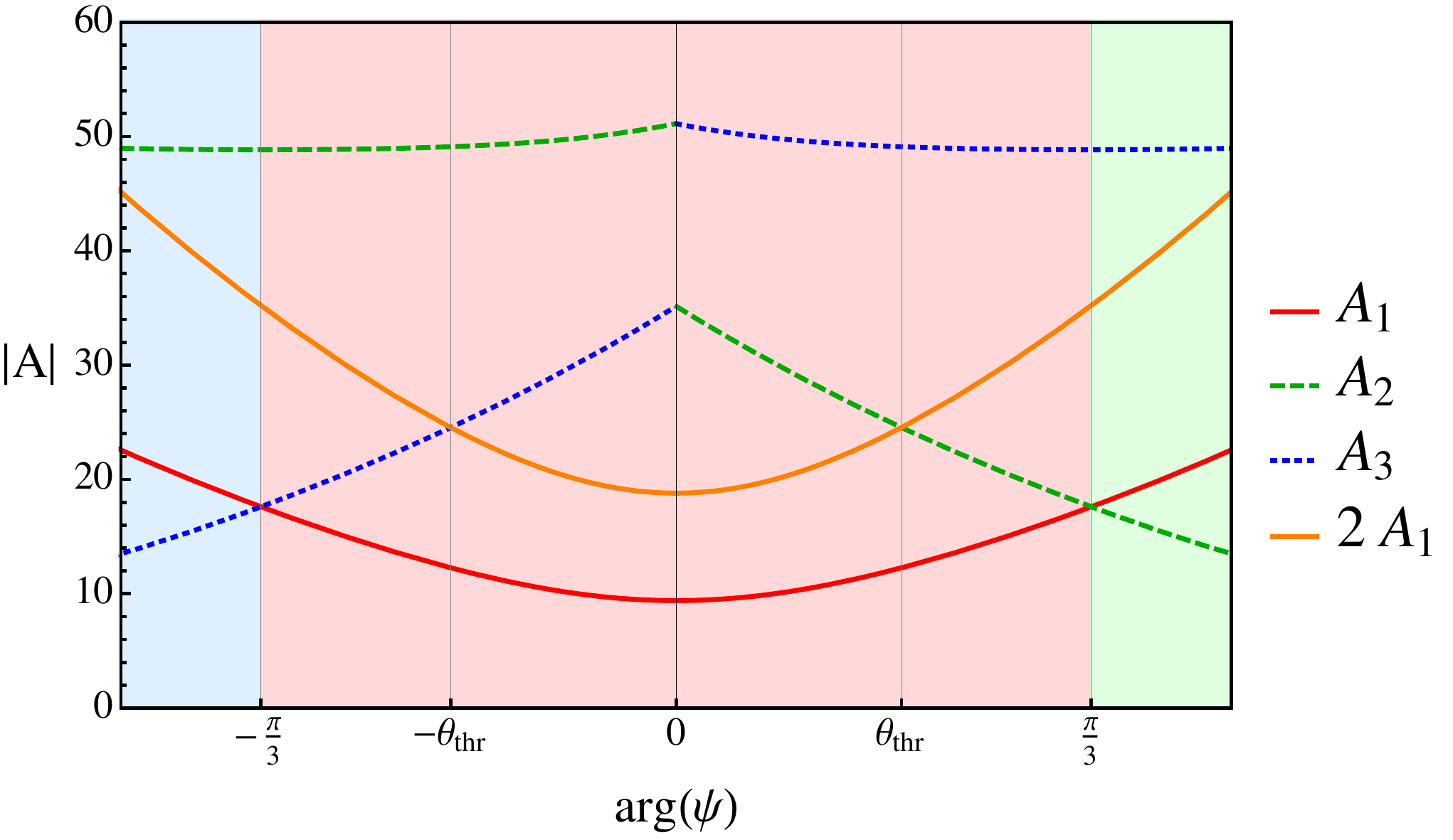}
  \captionsetup{width=5cm}
  \captionof{figure}{Dominant and subdominant instanton actions.\\}
  \label{fig:4}
\end{minipage}%
\begin{minipage}{.5\textwidth}
  \centering
  \includegraphics[height=.55\linewidth]{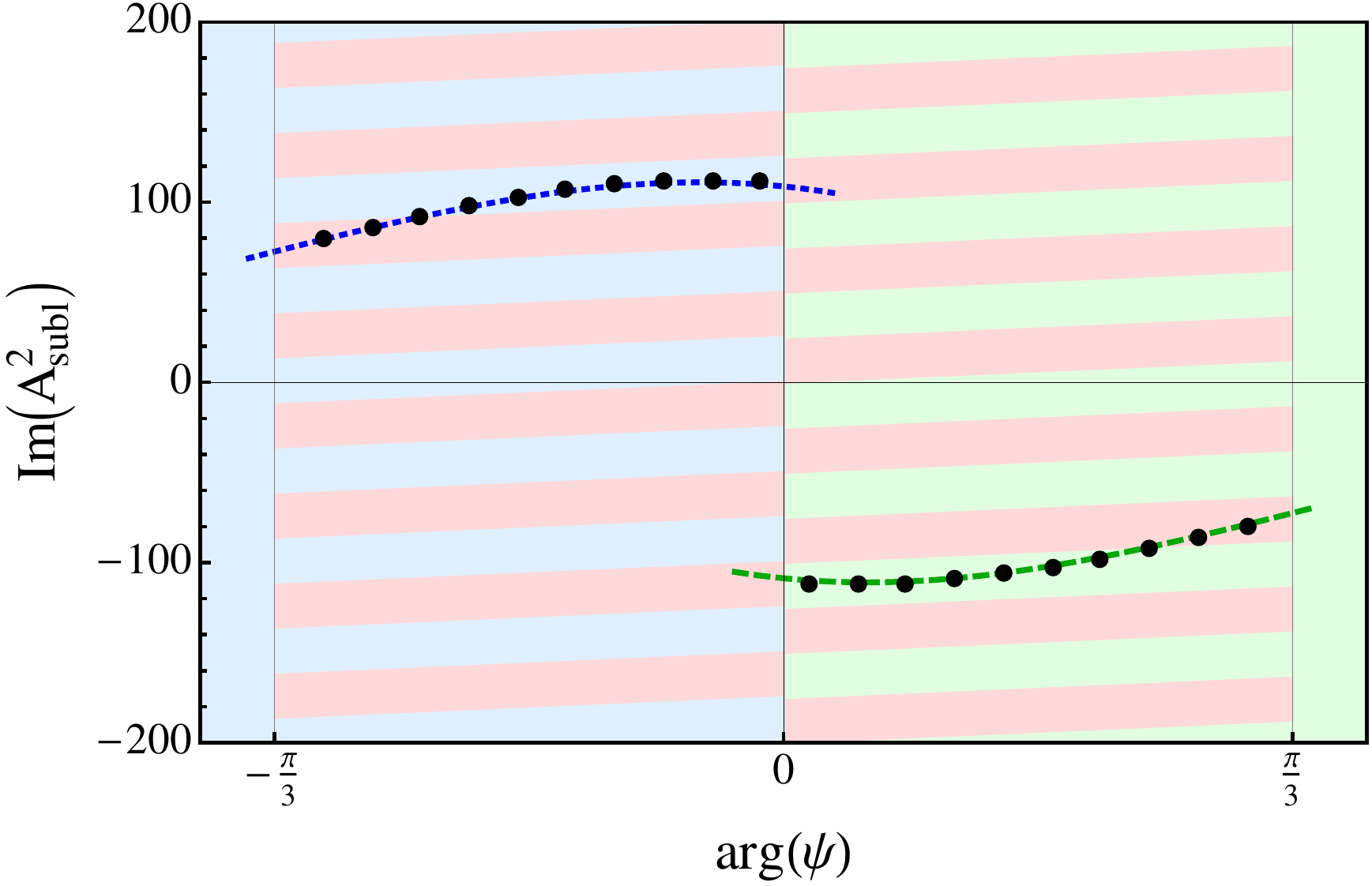}
  \captionsetup{width=5.1cm}
  \captionof{figure}{A crossover between two different subdominant one--instanton sectors.}
  \label{fig:5}
\end{minipage}
\end{figure}

As can already be seen from figure \ref{fig:2}, at different points in moduli space, different instanton actions have the smallest absolute value, meaning that different instanton actions determine the dominant large--order behavior of the perturbative coefficients. One can now try to Borel resum the contributions coming from a given instanton sector, and measure the contributions in the  large--order relations subleading to all of those. In fact, two things can happen: those subdominant instanton contributions can come from a {\em different} instanton, or they can come from {\em two--instanton} effects of the same type as the resummed one--instanton sector. Which of these is the case depends on the absolute values of the relevant instanton actions. This is plotted for a specific slice in moduli space in figure \ref{fig:4}. For example, for values of $\psi$ in the left of the figure, we see that the instanton action $A_3$ dominates, whereas the instanton action $A_1$ determines the subdominant large--order behavior of $F_g$. For values of $\psi$ near the middle of the plot, the $A_1$--sector dominates, and the subdominance comes from two--instanton effects corresponding to the same instanton.

\begin{figure}[t]
\centering
\begin{minipage}{.5\textwidth}
  \centering
  \includegraphics[height=.55\linewidth]{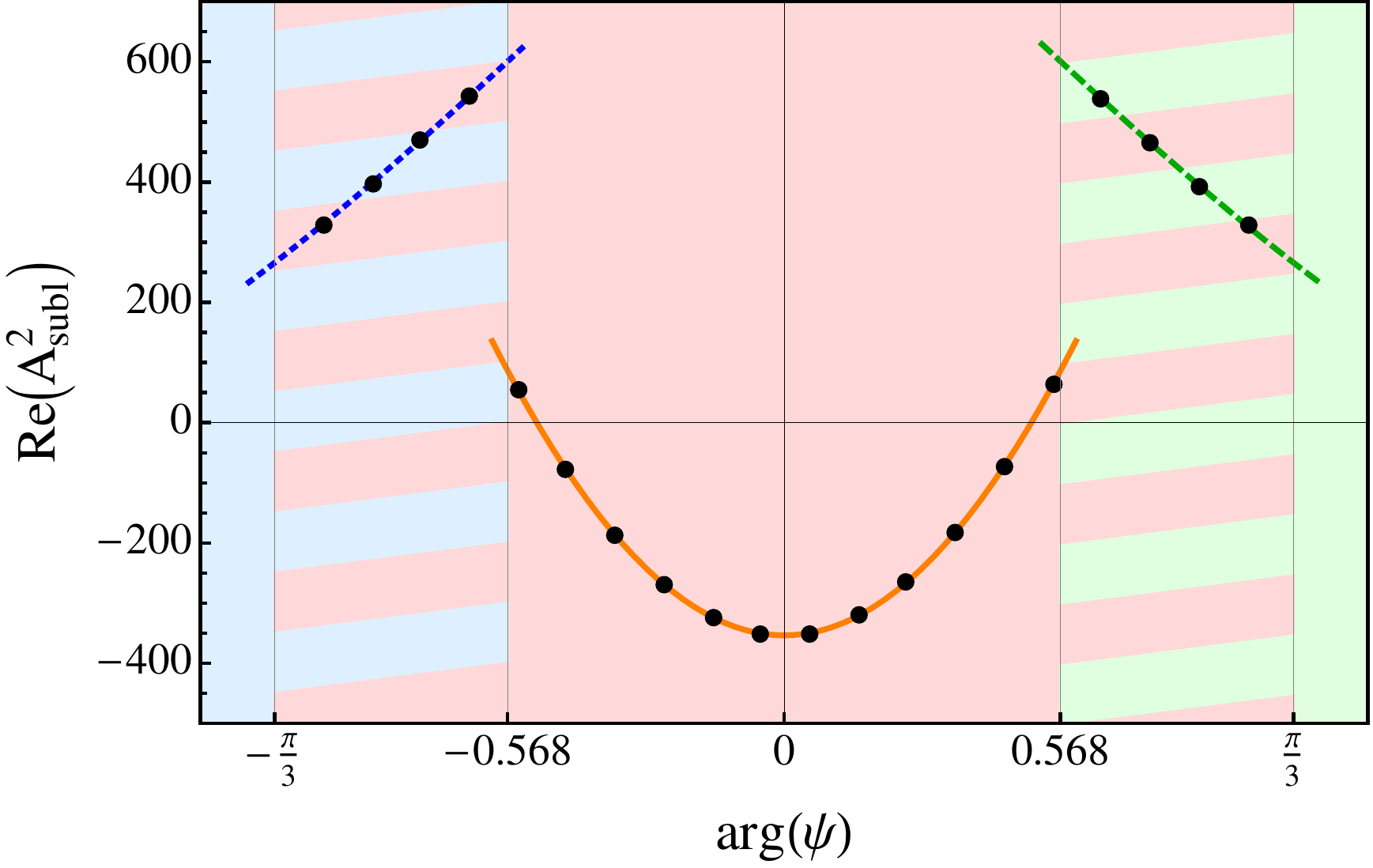}
  \captionsetup{width=6cm}
  \captionof{figure}{A crossover between subdominant one-- and two--instanton sectors.}
  \label{fig:6}
\end{minipage}%
\begin{minipage}{.5\textwidth}
  \centering
  \includegraphics[height=.55\linewidth]{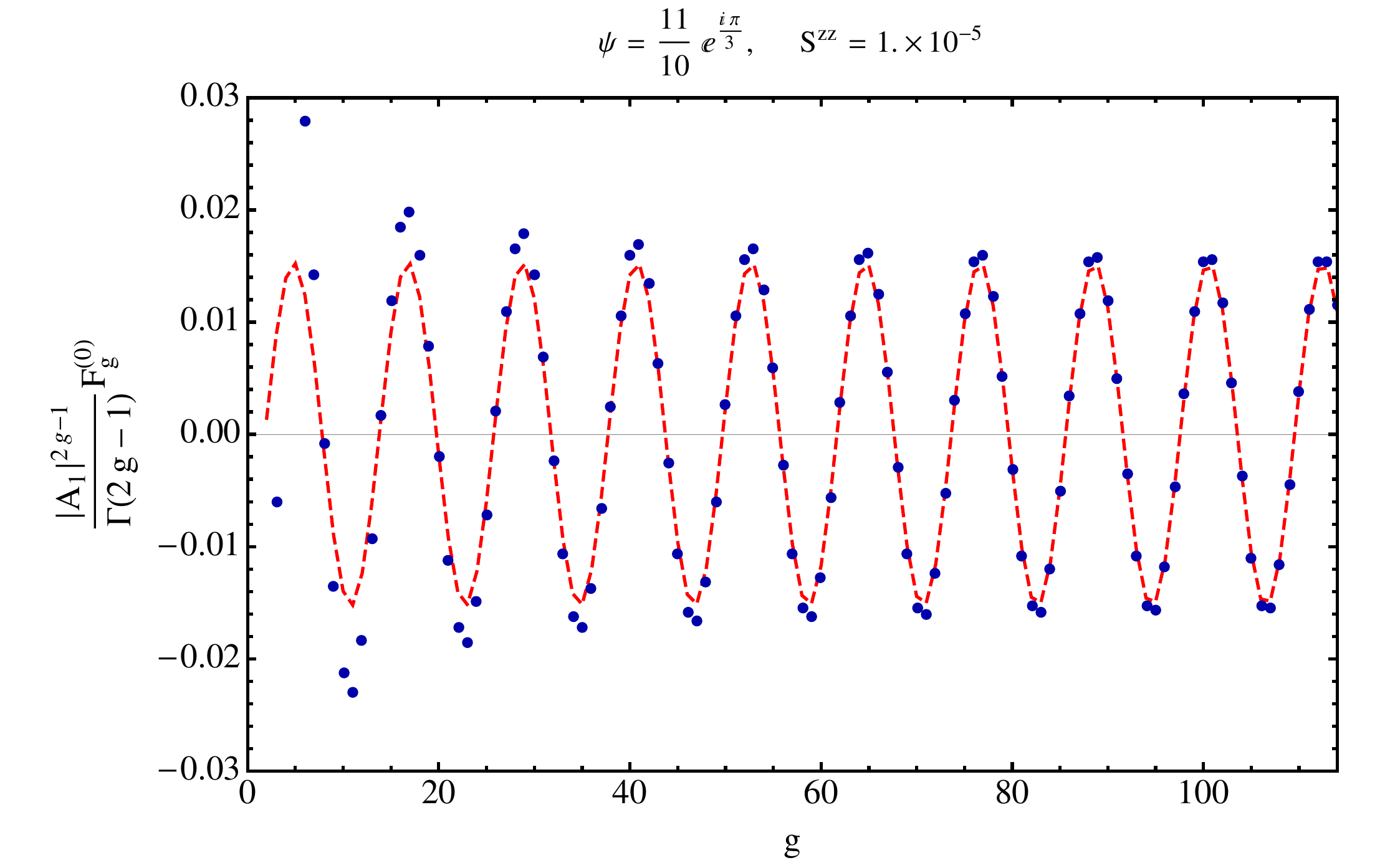}
  \captionsetup{width=5.5cm}
  \captionof{figure}{Oscillatory large--order behavior when two instanton actions dominate.}
  \label{fig:7}
\end{minipage}
\end{figure}

We have checked all of these predictions against large--order behavior, and found perfect agreement. For example, in figure \ref{fig:5} we zoom in to a region in moduli space where the subdominant behavior is caused by two different one--instanton sectors. After resumming the leading large--order behavior, we see that what remains (the dots in the figure) indeed perfectly matches the predictions from the analytic transseries solution. In particular, a jump in the large--order data occurs exactly where one would expect to see it. A similar plot in figure \ref{fig:6} shows a region where first a one--instanton sector is subdominant, then a two--instanton sector takes over, and finally a different one--instanton sector determines the subdominant behavior. Once again, the dots coming from the large--order perturbative data perfectly match the predictions from the resurgent transseries ansatz.

Finally, one may wonder what happens at points in moduli space where two instanton actions have exactly the same absolute value, so that each of the corresponding sectors gives an equal size contribution to the large--order behavior. It is not too hard to show that at those points, resurgence predicts that the large--order behavior of the $F_g$ obtains an oscillatory component. In figure \ref{fig:7}, we isolate the expected oscillatory behavior from the transseries prediction (the continuous line) and see that for large $g$ (plotted horizontally), the data again nicely matches the prediction.

\section{Conclusion and outlook}
\label{sec:conclusion}
All of the results in the previous section (and many more reported in \cite{Santamaria:2013rua,Couso-Santamaria:2014iia}) seem to indicate that the nonperturbative topological string free energy is indeed a resurgent function, and that therefore, resurgent transseries techniques based on the holomorphic anomaly equation can be used to give a proper nonperturbative completion of the topological string. While these results are very encouraging, a number of open questions remain, of which we mention the two most pressing ones:
\begin{itemize}
 \item
  Even though one can now {\em calculate} nonperturbative contributions to the topological string free energy, it would be nice to have a more physical {\em interpretation} of the effects that these contributions describe. In certain examples, through dualities, these effects turn out to match D--brane or tunneling effects, but a generic description from a purely topological string point of view is still missing.
 \item
  A crucial technical observation is that the holomorphic anomaly equation (\ref{eq:haint}) is a differential equation in $t$, and {\em not} in the transseries variable $g_s$. As a result (due to the lack of a so--called ``bridge equation'', as explained in \cite{Santamaria:2013rua}), one cannot fully derive the large--order relations and the transseries structure from first principles, but one has to resort to information obtained from either dual desciptions or large--order analysis. At the level discussed in this paper, this is not an issue, but on a more fundamental level it is. For example, one cannot a priori determine the number of different instanton actions, and once one goes to higher than leading subdominant contributions, it is no longer fully clear which transseries coefficients determine the large--order behavior. In fact, we cannot say with certainty whether the topological free energy is a so--called ``simple resurgent function'' (meaning that its Borel transform only has simple poles and logarithmic branch cuts) or whether it is of some more complicated type. It would be good to have a better mathematical handle on these issues.
\end{itemize}
Apart from these fundamental open questions, it would be good to study more examples --- for example, models with more moduli, or models with different (non--conifold) types of special points in moduli space --- in order to develop further the technical tools presented here. Undoubtedly, this will shed a lot of new light on the so far mysterious issue of the nonperturbative definition of the topological string.

\bibliographystyle{amsplain}

\begin{thebibliography}{99}
\bibitem{Santamaria:2013rua} 
  R.~C.~Santamar\'ia, J.~D.~Edelstein, R.~Schiappa and M.~Vonk,
  ``Resurgent Transseries and the Holomorphic Anomaly,''
  arXiv:1308.1695 [hep-th].
\bibitem{Couso-Santamaria:2014iia} 
  \bysame
  ``Resurgent Transseries and the Holomorphic Anomaly: Nonperturbative Closed Strings in Local $\mathbb{C}\mathbb{P}^2$,''
  arXiv:1407.4821 [hep-th].
\bibitem{Aniceto:2011nu} 
  I.~Aniceto, R.~Schiappa and M.~Vonk,
  ``The Resurgence of Instantons in String Theory,''
  Commun.\ Num.\ Theor.\ Phys.\  {\bf 6}, 339 (2012)
  [arXiv:1106.5922 [hep-th]].
\bibitem{Aniceto:2013fka} 
  I.~Aniceto and R.~Schiappa,
  ``Nonperturbative Ambiguities and the Reality of Resurgent Transseries,''
  arXiv:1308.1115 [hep-th].
\bibitem{Ecalle:1981}
  J.~\'Ecalle,
  ``Les Fonctions R\'esurgentes,''
  Pr\'epub.\ Math.\ Univ.\ Paris-Sud {\bf 81-05} (1981), {\bf 81-06} (1981), {\bf 85-05} (1985)
\bibitem{Hori:2003ic} 
  K.~Hori, S.~Katz, A.~Klemm, R.~Pandharipande, R.~Thomas, C.~Vafa, R.~Vakil and E.~Zaslow,
  ``Mirror symmetry,''
  (Clay mathematics monographs. 1)
\bibitem{Bershadsky:1993ta} 
  M.~Bershadsky, S.~Cecotti, H.~Ooguri and C.~Vafa,
  ``Holomorphic anomalies in topological field theories,''
  Nucl.\ Phys.\ B {\bf 405}, 279 (1993)
  [hep-th/9302103].
\bibitem{Bershadsky:1993cx} 
  \bysame
  ``Kodaira-Spencer theory of gravity and exact results for quantum string amplitudes,''
  Commun.\ Math.\ Phys.\  {\bf 165}, 311 (1994)
  [hep-th/9309140].
\bibitem{Haghighat:2008gw} 
  B.~Haghighat, A.~Klemm and M.~Rauch,
  ``Integrability of the holomorphic anomaly equations,''
  JHEP {\bf 0810}, 097 (2008)
  [arXiv:0809.1674 [hep-th]].
\bibitem{Drukker:2011zy} 
  N.~Drukker, M.~Marino and P.~Putrov,
  ``Nonperturbative aspects of ABJM theory,''
  JHEP {\bf 1111}, 141 (2011)
  [arXiv:1103.4844 [hep-th]].
\bibitem{Pasquetti:2009jg} 
  S.~Pasquetti and R.~Schiappa,
  ``Borel and Stokes Nonperturbative Phenomena in Topological String Theory and c=1 Matrix Models,''
  Annales Henri Poincare {\bf 11}, 351 (2010)
  [arXiv:0907.4082 [hep-th]].
\end{thebibliography}

\end{document}